\documentclass[aps,prd,english,showpacs,11pt]{revtex4-1}
\usepackage[latin1]{inputenc}
\usepackage{ucs}
\usepackage{amsmath}
\usepackage{amsfonts}
\usepackage{amssymb}
\usepackage{graphicx}
\usepackage{psfrag}
\usepackage{picinpar}
\usepackage[]{hyperref}

\begin{document}
\title{Domain Wall Model in the Galactic Bose-Einstein Condensate Halo}

\author{J. C. C. de Souza} 
\email{jose.souza@ufabc.edu.br}
\author{M. O. C. Pires}
\email{marcelo.pires@ufabc.edu.br}

\affiliation{Centro de Ci\^{e}ncias Naturais e Humanas, Universidade
  Federal do ABC, Rua Santa Ad\'elia 166, 09210-170, Santo Andr\'{e}, SP,
  Brazil}

\begin{abstract}
We assume that the galactic dark matter halo, considered composed of
an axionlike particles Bose-Einstein condensate \cite{pir12}, can
present topological defects, namely domain walls, arising as
the dark soliton solution for the Gross-Pitaevskii equation in a
self-graviting potential. We investigate the influence that such
substructures would have in the gravitational interactions within a
galaxy. We find that, for the simple domain wall model proposed, the
effects are too small to be identified, either by means of a local
measurement of the gradient of the gravitational field or by analysing
galaxy rotation curves. In the first case, the gradient of the
gravitational field in the vicinity of the domain wall would be
$10^{-31}\; (m/s^2)/m$. In the second case, the ratio of the
tangential velocity correction of a star due to the presence of the
domain wall to the velocity in the spherical symmetric case would be
$10^{-8}$.

\end{abstract}


\maketitle

\section{Introduction}

The existence of a mysterious kind of matter, rather different from
the usual barionic matter, presents itself as a great challenge for
modern Physics. This so-called dark matter corresponds to almost 23\%
of the energy density of the Universe \cite{wmap} and can amount to
approximately 90\% of the total mass in galaxies.

Recently, it has been proposed that this type of matter can be
composed of some kind of weakly interacting bosons
\cite{jaeckel}. When these bosons are spinless they can be identified
with axions, hypothetical particles proposed in the context of
Peccei-Quinn models \cite{peccei}. On the other hand, in the case of
sub-eV spin-1 particles, they are called hidden bosons or hidden
photons \cite{arias, nel11}. Axions and hidden bosons form a class of
particles known as WISPs (Weakly Interacting Slim Particles), due to
their diminute masses.

In the last few years, the possibility that the dark matter content of
galaxies is in the form of a self-graviting Bose-Einstein condensate
(BEC) has been considered. Using this approach, the authors in
\cite{boh07} were able to relate the mass and the scattering length of
an axionlike particle with the radius of the galactic dark matter
halo. By proposing a new density profile based in the BEC features
they could construct rotation curves that fit well a sample of
galaxies.

Using the same initial hypothesis, and extending it to spin-1
particles, the authors in \cite{pir12} showed that the mass of the
WISP's is constrained by galaxy radii data to the range
$10^{-6}-10^{-4} \; eV$. 

The next natural step in the identification of the dark matter halo
with a BEC is to study the possible presence of substructures. BEC's
can present a number of different substructures, called topological
defects, such as vortexes, domain walls, monopoles and textures.

The existence of these substructures is verified in laboratory
experiments, for ultra-cold alkali atom gases trapped in a magnetic
optical potential, and their features are well studied under these
circumstances \cite{bur99,sai07,abd05}. They are observed as a small
region (much smaller than the size of the condensate) of null mass
density in the gas. Mathematically, these defects have origin in zeros
of the condensate wave-function, stressing the quantum nature of these
phenomena in the gas.

Topological defects have also been studied in the framework of
cosmology and gravitation \cite{bra98}, in which they have notable
differences from the condensed matter ones. For instance, they can be
massive and carry a large amount of energy. Recently, the authors in
\cite{pos13} have suggested an experiment to detect a massive axionic
domain wall via magnetic interaction in the context of field theory.

To the authors knowledge, topological defects for a self-graviting BEC
have never been proposed before. Our goal in the present paper is to
explore the possibility that the galactic BEC is endowed with a domain
wall, a finite region in space where the density vanishes. We are
interested in the effect that such a structure can have on the
galactic dynamics (specially on rotation curves), and if it is
possible to detect a local domain wall by means of gravitation
interaction effects. We restrict ourselves to the axionlike particle
case.

\begin{figure}[!htp]
\begin{center}
\psfrag{z0}{{\tiny $z_0$}}
\psfrag{spiral galaxy}{{\tiny galactic disk}}
\includegraphics[scale=0.4]{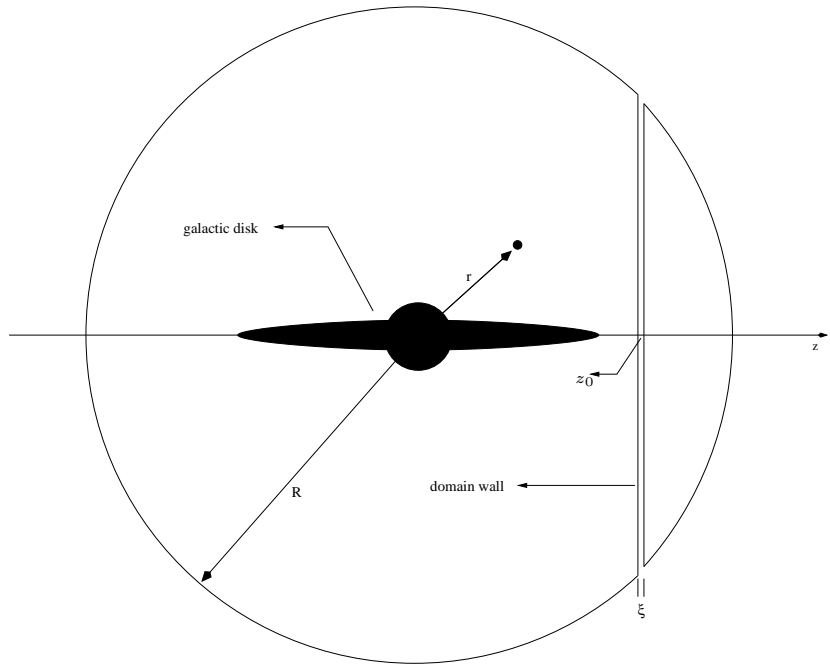}
\caption{Schematic view of the coordinate system used in this
  paper. $R$ is the galaxy halo radius. The domain wall of width $\xi$
  is in a position $z_0$, and is perpendicular to the galactic plane,
  chosen to be the location of the $z$ axis.}
\label{scheme2}

\end{center}
\end{figure}

This paper is organized as follows. In section \ref{dark} we perform
the derivation of the density functions for the halo endowed with a
domain wall, and estimate its width. In section \ref{local}, we give
an estimate of the gradient of the halo gravitational field in the
vicinity of a domain wall. In section \ref{rotation} we derive a
correction term for the rotation curves of stars in spiral galaxies
taking into account the influence of a domain wall perpendicular to
the galactic disk plane (as depicted in figure \ref{scheme2}). Section
\ref{conclusions} shows our final remarks.

\section{Dark solitons}\label{dark}

The zero-temperature mean field energy of a weakly interacting BEC
confined in a self-graviting potential, $V$, is given by \cite{pir12}
\begin{eqnarray}
E=\int d^{3}{\bf
  r}\left[\psi^*\left(-\frac{\hbar^2}{2m}\nabla^2+V\right)\psi\right]+\frac{4\pi\hbar^2a}{m}|\psi|^4
\end{eqnarray}
where $m$ is the mass of the particle composing the condensate, $a$ is
the $s$-wave scattering length and $\psi(\vec{r})$ is the
condensate wave function, satisfying $\int d^{3}{\bf r}|\psi|^2=N$
with $N$ being the total number of particles.

The mean field dynamics of the system is described by the
Gross-Pitaevskii (GP) equation
\begin{eqnarray}\label{GPE}
i\hbar \frac{\partial \psi({\bf r},t)}{\partial
  t}=\left(-\frac{\hbar^2}{2m}\nabla^2+V({\bf r})+g|\psi({\bf
  r},t)|^2\right)\psi({\bf r},t)\;,
\label{gpe}
\end{eqnarray}
where $g=4\pi\hbar^2a/m$.

Some of the solutions of equation (\ref{GPE}) may be quantized
vortexes or dark solitons. These functions are topological defects in
scalar BECs, in which the density vanishes due to the topological
constraint on the phase of the wave function.

In order to investigate the topological defect that corresponds to the
domain wall that can appear perpendicularly to the $z$ direction, we
coupled the topological defect with the ground state of the galactic
condensate in the Thomas-Fermi approximation \cite{boh07} in the form
\begin{eqnarray}
\psi({\bf r},t)\equiv\psi_{TF}(x,y,z)\phi(z,t)\;,
\label{ans}
\end{eqnarray}
where $\psi_{\small{TF}}(x,y,z) \equiv \psi_{\small{TF}}(r)$ is the Thomas-Fermi
solution for the GP equation (with $r^2= x^2+y^2+z²$),
\begin{eqnarray}
\psi_{TF}(r)=\begin{cases} \sqrt{\rho_0\frac{\sin
    kr}{kr}}\quad\mbox{for}\quad r\le R \\ 0\quad \mbox{for} \quad
r>R \end{cases}
\end{eqnarray}
with $k=\sqrt{Gm^3/\hbar^2a}$, $R=\pi/k$ is the condensate radius and
$\rho_0$ is the central number density of the condensate. $\phi(z,t)$
corresponds to the topological defect solution.

We are interested in characterizing the defect by its position $z$,
hence we eliminate the Thomas-Fermi solution in GP equation, as well as its 
dependence on $x$ and $y$ coordinates, by multiplying (\ref{GPE}) by
$\psi_{TF}^{*}$ and integrating it in these coordinates, obtaining
\begin{eqnarray}
i\hbar \frac{\partial \phi(z,t)}{\partial
  t}=\left(-\frac{\hbar^2}{2m}\frac{\partial^2}{\partial
  z^2}-\eta(z)\frac{\partial}{\partial
  z}+V+g(z)|\phi(z,t)|^2\right)\phi(z,t)\; ,
\label{gpe1d}
\end{eqnarray}
where $\eta(z)=-\frac{\hbar k}{2m}\frac{\sin(kz)}{(1+\cos(kz))}$ is an
extremely small factor and the term it couples to can be neglected.

The effective interaction parameter, $g(z)=g\rho_0f(kz)/2$, is
proportional to the central density and the form factor
\begin{eqnarray}
f(x)=\frac{\ln\left(\frac{1}{x}\right)+\int_{2\pi x}^{2 \pi}\frac{\cos
    (t)}{t}dt}{1+\cos (\pi x)},
\end{eqnarray}
where $x=z/R$.
\begin{figure}[!htp]
\begin{center}

\includegraphics[scale=1.0]{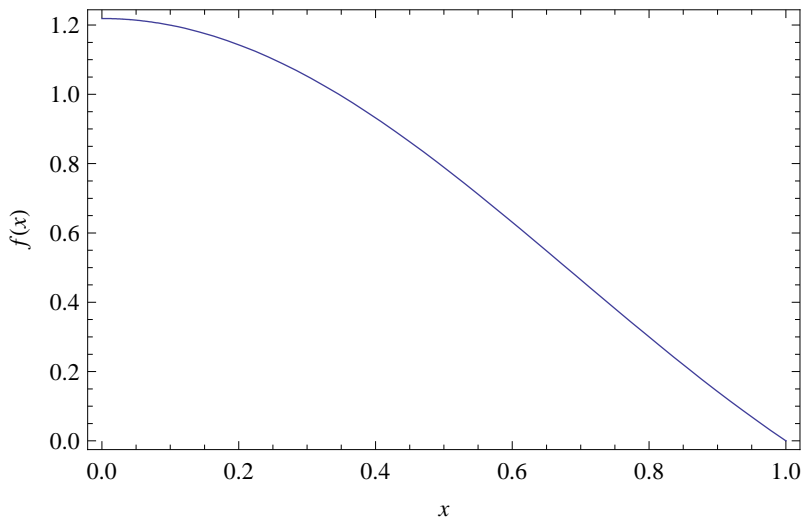}
\caption{Form factor of the condensate.}
\label{f_x}

\end{center}
\end{figure}

We assume that the width of the topological defect is much smaller than the size
of the condensate. In this situation we can consider the particle
number density and the effective interaction parameter as almost constants in
the defect vicinity. As the self-graviting potential obeys the Poisson
equation, we can approximate the potential as $V({\bf r})\approx V_0$
and, therefore, we substitute
\begin{eqnarray}
\phi(z,t)=u(z,t)e^{-iV_0t/\hbar}
\end{eqnarray}
in equation (\ref{gpe1d}) to obtain the one-dimensional Gross-Pitaevskii equation for $u(z,t)$
\begin{eqnarray}
i\hbar \frac{\partial u(z,t)}{\partial
  t}=\left(-\frac{\hbar^2}{2m}\frac{\partial^2}{\partial
  z^2}+g'|u(z,t)|^2\right)u(z,t),
\label{gpe1}
\end{eqnarray}
where $g'$ is supposed be locally constant. Equation
(\ref{gpe1}) has the solution
\begin{eqnarray}
u(z,t)=e^{-i\mu t/\hbar}\tanh\left(\frac{z-z_0}{\sqrt{2}\xi}\right),
\end{eqnarray}
where $\mu=g'=g(z_0)$. This solution is called a planar dark soliton,
describing a domain wall at $z=z_0$, since the density vanishes at
that point.

The quantity $\xi$, called healing length, is related to the width of
the domain wall. It is possible to show that it is a function of the
parameters of the condensate in the form 
\begin{eqnarray}\label{csi}
\xi=\frac{1}{\sqrt{4\pi \rho_0 af(kz)}}.
\end{eqnarray}

\begin{figure}[!htp]
\begin{center}

\includegraphics[scale=1.0]{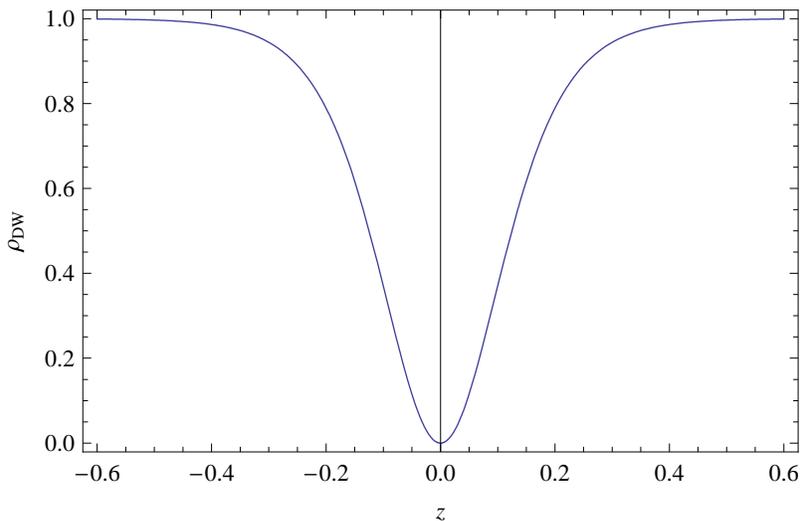}
\caption{Density function for the domain wall located at the origin of
  the coordinate system and with a healing length of 0.1 (in arbitrary
  units of length).}
\label{rho_wall}

\end{center}
\end{figure}

The density function for the domain wall $\rho_{DW}=|u(z,t)|^{2}$ is
shown in figure \ref{rho_wall}.

\begin{figure}[!htp]
\begin{center}

\includegraphics[scale=1.0]{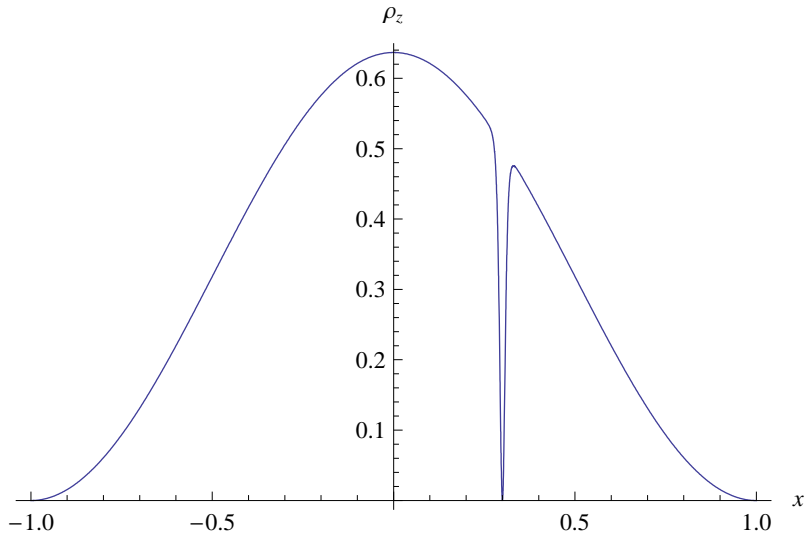}
\caption{Density function (in units of $k^2/(2 \rho_0)$) for the
  condensate endowed with a domain wall. The width of the domain wall
  has been made large in order to facilitate visualization.}
\label{rho_total}

\end{center}
\end{figure}

The density function 
\begin{equation} 
\rho_z=\int |\psi({\bf r},t)|^2 dx dy
\end{equation} 
for the condensate with a domain wall is depicted in figure
\ref{rho_total}.

Local dark matter density measurements indicate a value of $0.4\,
GeV/cm^{3}$. Using that information and the masses and scattering
lengths for an axionic dark matter halo as suggested in \cite{pir12}, we
can infer the order of magnitude for the healing length of the
domain wall. The results are shown in table \ref{tabela-1}. 
  
\begin{table}\centering
\caption{Values for the healing length $\xi$ using masses and
  scattering lengths obtained in \cite{pir12}.}\label{tabela-1}
\begin{tabular}{ |c|c|c| }
  \hline                        
  m (eV) & a (fm)& $\xi$ (m)  \\
  \hline
$10^{-6}$ & $10^{-14}$ & $10^{3}$  \\
  $10^{-5}$ & $10^{-11}$ & $10^{2}$  \\
  $10^{-4}$ & $10^{-8}$ & 10\\
\hline  
\end{tabular}
\end{table}

We can see that the healing length decreases very quickly with the
particle's mass, becoming negligible for larger masses. The value for
a mass of $10^{-4}\; eV$ is already beyond any physical significance
at galactic scales.

\section{Gravitational effect in the vicinity of the domain wall}\label{local}

We proceed now to the calculation of the effect of a domain wall
located on the galaxy disk, more specifically crossing Earth's
position, on a test body (e. g., a satellite).

In the presence of a domain wall, the total density distribution is
not symmetrical, then the gravitational effects are distinct from the
case of the halo density without the domain wall. We intent to
estimate this difference by analysing the movement of a massive test
body crossing the domain wall.

The gravitational field on
the body is given by the solution of the equation
\begin{eqnarray}
\nabla\cdot \vec{g}=-4\pi G\varrho,
\end{eqnarray}
where $\varrho$ is the mass density which can be related to the
wave function by
\begin{eqnarray}
\varrho(x,y,z)=m|\psi({\bf r},t)|^2\; .
\end{eqnarray}

The gravitational effect in the test body will be maximized if
this body is moving along the z-axis and between the border and the
center of the domain wall. By the symmetry of the spatial
configuration, the gravitational field has only a z-direction
component. In this case, the equation to be solved is
\begin{eqnarray}
\frac{\partial }{\partial z}g_z(z)= -4\pi Gm\rho_0\frac{\sin
  (kz)}{kz}\tanh^2\left(\frac{z-z_0}{\sqrt{2}\xi}\right).
\end{eqnarray}
The difference in the gravitational field is given by
\begin{eqnarray}
g_z(z_0)-g_z(z_0-\xi/2)
\approx -4\pi
Gm\rho_0\xi\frac{\sin (\pi x_0)}{\pi
  x_0}\left(\sqrt{2}\tanh\left(\frac{1}{2\sqrt{2}}\right)-1\right),
\end{eqnarray}
where $x_0=z_0/R$ is the domain wall position relative to the galaxy
radius. 

For the Sun's relative position, $x_0\sim 0.5$ and the gravitational
effect exerted in the massive body crossing the topological defect is
of the order of $10^{-28}\; m/s^{2}$, for a healing length of the
order of $10^{3}\; m$ and the gradient of the gravitational field is
$10^{-31}\;(m/s^2)/m$. Because the Earth's movement (along with the Sun)
in the galaxy has a velocity of $\sim 10^{5}\; m/s$, this effect in the
vicinity of our planet could only be detected by an experiment with a
precision greater than $10^{-32}\; m$, which is far beyond present day
technological capability.

\section{Tangential velocity correction}\label{rotation}

When the domain wall is present, the gravitational field presents
tangential components.  However, the projection of the gravitational
field in the tangential direction is much smaller than the projection
in the radial direction even near the domain wall. Then, we can
neglect the tangential components and assume that the gravitational
field is radial and given by
\begin{eqnarray}
\frac{1}{r^2}\frac{\partial }{\partial r}r^2g_r(r)= -4\pi
G\varrho(r,\theta,\phi),
\label{gfield}
\end{eqnarray}
where
\begin{eqnarray}
\varrho(r,\theta,\phi)=m\rho_0\frac{\sin
  (kr)}{kr}\tanh^2\left(\frac{r\cos(\theta)-z_0}{\sqrt{2}\xi}\right).
\end{eqnarray}

Using Gauss theorem in the equation (\ref{gfield}), we obtain
\begin{eqnarray}
g_r(r)=-\frac{GM_{DM}(r)}{r^2},
\label{radialfield}
\end{eqnarray}
where the mass profile of the dark condensate galactic halo is,
\begin{eqnarray}
M_{DM}(r)=\int_{\mathcal{V}} \varrho (r,\theta,\phi)d^3r,
\end{eqnarray}
with $\mathcal{V}$ the volume of a sphere with radius $r$.

Equation (\ref{radialfield}) allows to represent the tangential velocity
$v^2_{tg}(r)=rg_r(r)$ of a test particle moving in the halo as
\begin{equation}\label{vsquared}
v^{2}_{tg}(r)=v^{2}_{ss}(r)-v^{2}_{corr}(r)\; ,
\end{equation}
where
\begin{eqnarray}
v^2_{ss}(r)=\frac{4\pi Gm\rho_0}{k^2}\left(\frac{\sin (kr)}{kr}-\cos
(kr)\right)
\end{eqnarray}
is the squared tangential velocity for the spherically symmetric case
(already obtained in \cite{boh07})
and
\begin{eqnarray}
v^{2}_{corr}(r)=\frac{4\pi
  Gm\rho_0}{k^2}\left(\sqrt{2}\pi\frac{\xi}{R} \Theta(r-z_0)\frac{\cos
  (kz_0)-\cos (kr)}{kr}\right)
\end{eqnarray}
is the correction in the squared velocity due to the presence of the domain
wall. $\Theta(x)$ is the Heaviside step function.

$v^{2}_{corr}$ is proportional to $\xi/R\ll1$, then the correction is
maximal when the wall is located near the center of the
galaxy. As the domain wall width is always many orders of magnitude
smaller than the radius of the halo, this correction is also small.

\begin{figure}[!htp]
\begin{center}

\includegraphics[scale=1.0]{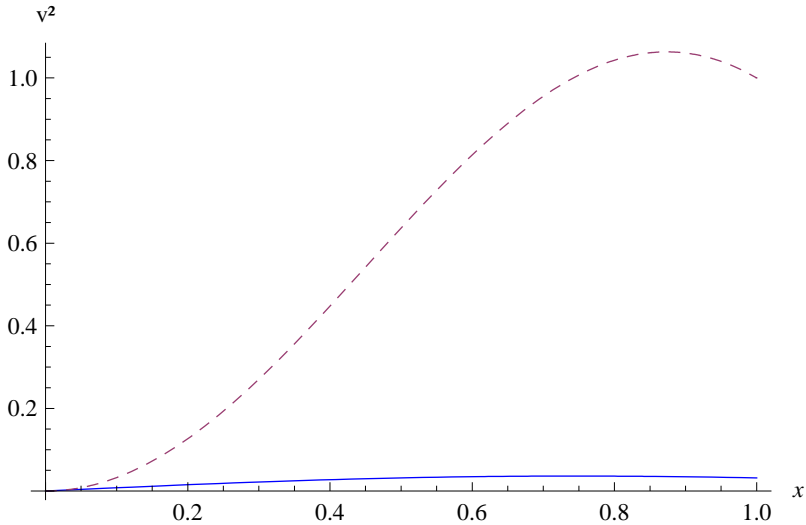}
\caption{Tangential velocities (in $km^{2}/s^{2}$) for the BEC
  dark matter halo (dashed line) and the domain wall (solid line). The
  wall's relative width $\xi/R$ has been chosen as 0.05 to allow easy
  visualization of both curves. It is possible to see that the
  diference between the curves is very large. Here $x=r/R$.}
\label{v2}

\end{center}
\end{figure}

In figure \ref{v2} both terms of (\ref{vsquared}) are shown, to
stress the difference in magnitude they present.

\begin{figure}[!htp]
\begin{center}

\includegraphics[scale=1.0]{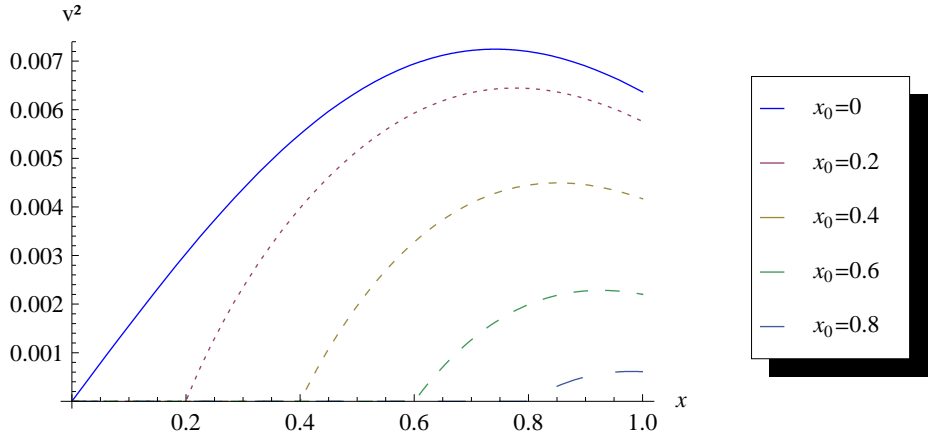}
\caption{Tangential velocity correction due to the influence of the
  domain wall with a relative width of 0.01. The curves are related to
  walls located in increasing relative distances $x_0$ from the center
  of the galaxy.}
\label{v2_mult}

\end{center}
\end{figure}

With the addition of a barionic matter term (after choosing an
appropriated barionic matter density profile), equation
(\ref{vsquared}) can represent a rotation curve for stars in a spiral
galaxy. The term $v^{2}_{ss}$ had already been obtained in
\cite{boh07}, and it was found to fit observed rotation curves for a
number of galaxies. Our correction, $v^{2}_{corr}$, because of its
small magnitude, cannot be detected in these types of rotation
curves. As the factor $\xi/R$, for typical galaxies, would amount to
about $10^{-16}$, the ratio between the correction and the tangential
velocity for the spherical symmetric case would be
$v_{corr}/v_{ss}\sim 10^{-8}$.

The small efect that a domain wall would have in the galactic dynamics
can be explained when we take in consideration the mass that could
fill a disk the same size as the domain wall in a galaxy similar to
the Milky Way ($R\approx 10\; kpc$). This mass would amount roughly to
$10^{22}\; kg$, or about the mass of the Moon.

\section{Conclusions}\label{conclusions}

By assuming that the dark matter halo in galaxies is composed of a
condensate of bosonic particles (with axionlike properties), as
previous works hypothesized, we were able to model one type of
substructure in the halo, in the form of a topological defect known as
domain wall, derived from a dark soliton solution for the
Gross-Pitaevskii equation in a self-graviting potential. 


Because other types of topological defects (such as vortexes,
monopoles, textures and ring solitons) would occupy a smaller volume
in the halo, and therefore would have a smaller influence on the total
dark matter density, we decided to restrict to the study of domain
walls. Even in this case, the magnitude of the effects are too small
to be subject to detection by present methods, at least for the
choice of parameters (mainly the healing length $\xi$, which depends
on the mass and the scattering length of the axionlike dark matter
particle estimated in \cite{pir12}) and simplifications we have made
here. For the local gravitational interaction, we have a gradient in
the field of the order of $10^{-31}\; (m/s^2)/m$. The correction
factor on the velocity rotation curves for stars in spiral galaxies is
typically of the order of $10^{-8}$.

However, there may exist some kind of cumulative effect that renders
the influence of domain walls considerable in a system with a larger
number of topological defects or a greater dark matter density. They
also may be important in other phases of galaxy evolution.

The main result of this work is the implementation of a methodology
for the inclusion of topological defects in a quantum gas dark matter
halo. The sequence of this study implies, for example, the
determination of the dynamical and thermodynamical stability of the
domain wall, the extension of the method for a spin-1 particle
condensate (as suggested in \cite{pir12}), and the manifestation of
such defects in a fermionic quantum fluid, among other
possibilities. These issues will be the subject of future work.

\begin{acknowledgments}
 J. C. C. S. thanks CAPES (Coordena\c c\~ao de Aperfei\c coamento de
 Pessoal de N\'\i vel Superior) for financial support.
\end{acknowledgments}

\end{document}